\documentstyle [12pt,psfig]{article} 
\begin{document}
\hsize 15.5 cm
\vsize 22 cm
\centerline {\Large Mathematics, Brain Modelling \& Indian Concept of Mind$^*$}

\bigskip

\centerline {Bikas K. Chakrabarti}
\centerline {Saha Institute of Nuclear Physics}
\centerline {1/AF Bidhan Nagar, Kolkata 700064.}

\vskip 1cm

\noindent Abstract: We describe briefly the recent advances in 
understanding the distributed nature of computations in the
(neural) network structure of the brain.
 We discuss if such artificial networks will
be able to perform mathematics and natural sciences. The problem
of consciousness in such machines is addressed. Ancient Indian
ideas regarding mind-body relations and J. C. Bose's experimental
observations regarding the highly distributed computations in
the  plant body is discussed.

\bigskip
 
\noindent {\bf I. Mathematics and logic:}
\medskip

\noindent What is the nature of mathematical truths? Is mathematical
knowledge true a priori? Independent of experience and observation?
Does one have to verify mathematical truths in any (mathematical) 
laboratory? If not, what makes it true?

Philosphers and mathematicians have thought about it for a long time:
see e.g., Whitehead and Russell [1,2] for some discussions
on these thoughts by western thinkers (a comparative study of eastern
and Indian thoughts seems to be missing -- at least not known to this
author). Although several ideas had been developed over the ages, 
Whitehead and Russell proposed very forcefully, following Hilbert (1862
- 1943), the idea that mathematical statements are true because of
their internal consistency and as they do not convey anything new; they are
in fact tautologies. Mathematics
is just a condensed form of logic. Mathematical proof of a proposition is
just an elaboration of the proposition itself; nothing new is conveyed or
introduced by mathematical proof. That is why, there is no need to have
a laboratory checking the truth of a mathematical statement. Two plus
two is four because the concept of (the set of) four contains the
concept of (the sets of) two. That fifty minus fifty makes it zero
need not be checked by pushing all the fifty odd audience  out of this
lecture hall! It is true just like the truth of a statement 
``A bachelor does not have any wife"; one need not 
check if each individual  bachelor 
satisfies it or not. To prove the truthfulness of these statements, one 
just needs to look at the meaning of the words involved!

\bigskip

\noindent----------------------------------------------------

\noindent 
$^*$ Based on a lecture delivered 
at the Asiatic Society, Kolkata, on March 15, 2002
(Conf.  Mathematics \& Astronomy in Ancient India, Asiatic Society, March 2002).

\newpage 

If it is true that mathematics is just a condensed form of logic, it
is certainly formalizable. In that case, is the human brain unique
or even necessary for doing and developing mathematics? Or, can a 
machine do that? May not be today in a complete form, but in some future 
day? Mathematics is certainly necessary; but are human mind or the mind
of a mathematician absolutely necessary? May be, we do not have the 
machine or the computer yet to replace the mathematicians; but in the
future, can we dispense with them?

Although the debate still continues, it appears that a major part of
mathematics like arithmetic is already formalizable and indeed a computer 
can do it and do it better. Question arises, what about analytical
mathematics and geometry? Intuitionists like Kronecker (1823-1891),
Poincar\'e (1854-1912), Borel (1871-1956), Weyl (1885-1955) and others 
chiefly argued against such formalized view of mathematics on the ground
of its inappropriateness in analysis and geometry (see e.g., [2]). For
example, how does such formalization work when a function or a number is
expressed as an infinite series? When all the terms in the series, 
which are part of the function or the number, can not be enumerated, as
in many such series forwarded by Ramanujam (1887-1920)? Similar are the
cases of geometrical analysis. A celebrated demonstration of the 
problem came from G\"odel (1906 - 1978). Seizing upon paradoxical 
situations like ``The statement I am making now is untrue", G\"odel
was able to show that if a formalized mathematics, as proposed in 
Principia Mathematica is wide enough, then (a) the system is necessarily
incomplete in the sense that there exists a formula {\it F} of the 
system such that neither {\it F} nor its negation is derivable, and
(b) if the system is consistent, then no proof of its consistency is
possible which can be formalized within it (cf. [2,3]). Later works of
people like Feigenbaum  on the bifurcation route to chaos in 
some nonlinear maps and its universality class obtained using 
renormalization group technique (a physical or inductive 
principle, not a mathematical
one), or like Witten on the string-dynamical description of elementary
particles indicated the existence and need of (computational or
physical)  
laboratory of mathematics (cf. [4]).

\bigskip

\noindent {\bf II. Inductive logic and pattern recognition:}

\medskip

\noindent As discussed in the previous section, mathematics is thought 
primarily to be based on deductive logic. Physics or for that matter other 
natural sciences are based on inductive logic. Here, based on careful
observations and `inductions' from there, one formulates the basic 
statements or truths. From such inductive truths, one then deduces the
special statements or truths appropriate for specific situations in 
physical or natural sciences. This deductive (logic) part in natural
sciences is of course an integral part and naturally involves 
mathematics. Unlike most of mathematics (if not all) these inductive
truths are obtained from the observations in various `laboratories'.
Looking at the sun rising on the east every morning, we come to the
inductive truth ``The sun rises in the east every morning". This helps us
to predict {\it almost} certainly, that the sun will come up in the 
east tomorrow. But unlike the mathematical or deductive truths, this 
inductive truth and its predictions are provisional; an observation next
day might force us to update, refine or change the truth. Compared to this, the
prediction that two stars and two others in sky in the night tomorrow
will make four stars, is not provisional and one need not have to
check if it was valid yesterday!

The search and application of these inductive logic are admittedly the
hallmark of human brain; more specifically of the brains of great
scientists. Unlike mathematics, which is thought to be mostly formalizable
and based on deductive logic, the natural sciences are thought to be 
essentially based on inductive logic or truth. Unlike the major part
of mathematics therefore, which can be performed by a machine or
computer, natural science is expected to be 
essentially developed by and dependent on 
the human mind or brain!

Recently, the computer scientists have developed algorithms to `search
or recognise patterns' in seemingly unrelated situations or sequences.
These mechanical processes of pattern recognition are indeed similar
in spirit to the search of inductive truths by human mind or brain! For
example, by recognising the `abnormalities' in the regular pattern of 
blood flow through our veins, the physicians diagonose our illnesses. Such 
an expertise  of a medical doctor is of course much too rudimentary compared
to that of natural scientist recognising, for example, that the same pattern
is involved in the motion of the planets around the sun and the apple
falling on the ground from the tree! Nevertheless, in the extended 
language of the computer science, all these are pattern recognition problems;
some are much simpler compared to others. Basically, the problems of the
medical doctor and of Newton are the same; pattern recognition.
Indeed, the above mentioned simple pattern recognition of blood flow pulses
can often be performed these days by `expert computer algorithms' as well.
Slightly more complicated recognition problems like that of elementary musical
rhythms and of hand-writing etc are now analyzable using computers; several
algorithms are already developed to help solving such problems. The 
recently developed `associative memory' algorithms by Hopfield (in 
1984; see the
next section and ref. [5]) and of `learning'
by reorganising the connections through interactive minimization 
of errors by the multi-layer perceptrons,
originally deviced by Rosenblatt (in 1962; see next section and ref. [5]) are
indeed very encouraging. To some natural scientists, these are clear
indications (inductive truth?) that all our deductive and inductive
logics, and hence entire mathematics and the natural sciences, can {\it in
principle} be performed by machines: computers or artificial neural networks.
This is a very recent and highly exciting problem posed in the literature
of philosophy of science, often known as the `Strong Artificial Intelligence 
(AI)' hypothesis; see e.g., Crick [6] 
supporting the hypothesis and the excitement
and Penrose [3], Chalmers [7] et al, contradicting parts of it (essentially
arguing from the G\"odel's theorem). We will continue with the discussion
on this issue at the end of the next section.

\bigskip

\noindent {\bf III. Neural network modelling of the brain:}
 
\medskip

\noindent 
Although we differ from the animals in almost all the parts of 
our bodies, our essential
difference is rightly identified in our respective brains. 
We are known by our brains; a mathematician, a scientist or
an artist differ essentially in their respective brains. Although
even in the physiques of mine and of mathematicians-scholars like
Aryabhatta II (b. 476 A.D.) or Bhaskaracharyya II (b. 1114 A.D.)
there must have been a lot of differences, 
the essential difference, as we are all too aware, have been in
our brains! But what is the precise physical difference between
my brain, and that of Aryabhatta or for that matter that of a cow? We do not
know yet about all the differences, except for their  size and 
the structure.

We know today, at least in principle, the structure and working mechanisms
of almost all the parts of our body, except for the brain, and have in fact
developed some artificial supplements for them. They are used when parts
of our body fail to function normally. We know that the outer-retina part 
of our eye is like a camera, heart is like a pump, etc and we can be
provided with artificial supports like spectacles, pacemakers, etc to
supplement their partial failures. In case of the brain, however, we are 
still helpless, even if it fails minimally. Our interest in the brain
structure and function is therefore not just of epistemological, mathematical,
computational or physical curiosity or interest, medical support possiblity in
future is of extreme importance and can hardly be overemphasized.

During evolution, the animal bodies developed their brains to perform the
primary task of helping the body act according to the changes in the
environment: to analyse the signals received from the environment and to
respond accordingly. As the body surface receives the external
signals, each portion of it is mapped in the brain. In fact, as
the body surface grows with the body volume to the power 2/3, the brain mass
varies with 2/3rd power of the animal body mass; bigger brain of the
elephant is required for the control of the bigger body. By injecting 
coloured stains inside a dead brain, about hundred and fifteen years back,
 the spanish doctor Ramon y Cajal (nobel prize in Medicine in 1906) showed that
the brains are made up of many tiny cells, called since then neurons, which
are connected to each other through synaptic junctions seperated by
semipermeable membranes. We now know almost certainly that human brain 
contains about $ 10^{12}$ neurons and we are all born with them; Aryabhatta, 
Newton, Einstein, Tagore and myself were all born with more or less this
number of neuronal cells in the brain. This number does not differ
much within the species, but differs considerably from species to
species; e.g., the birds have about $10^{8}$ neurons. 
The physical structure of a neuron is indicated in Fig. 1. Each neuron is
an electrical device capable, in principle, of a very simple (electrical)
operation. It collects electrical pulses (of millivolt order) from $10^4$
to $10^6$ other neurons connected to it through its dendrites. These
pulses, collected over a synaptic period of a few milliseconds, are then
summed-up in the cell body. If the resultant sum exceeds a threshold voltage,
a few tens of millivolt order, the neuron fires and a millivolt order 
electrical pulse propagates (at a speed of few meters per second) through
the axon (cable). It  then  passes over to the other connected neurons
through the respective synaptic junctions. As noted by several 
neurophysiologists, including Hebbs (in 1949), these synaptic connections 
between the neurons develop with training and learning. We are not born
with all these electrical wirings (synaptic connections) among the components
(neurons), although a significant fraction of them seem indeed to be
determined by hereditary factors.
 Needless to mention here that according to this picture, I
differ from Aryabhatta in developing my inter-neuronal connections;
not in our brain size or neuron number. These synaptic connections may be 
both excitory (where a positive pulse flow accross it keeping its
phase unchanged) and inhibitory (where a positive pulse passes over to
the connected neuron as negative pulse, with changed phase). In fact, such
random  millivolt order $10^4$ to $10^6$ incoming pulses add
up to only about $10^{-2}$ or $10^{-1}$ volts in the cell body of a 
single neuron. For all excitory or all inhibitory connections, this sum
in a single cell would go to an extremely high value and cause the failure
of the cell. More importantly, as we will see later, the absence of
`frustration' (see e.g., [5]) in the cases of all excitory or all 
inhibitory connections would reduce enormously the brain memory capacity.
As may be noted from Fig. 2, these synaptic connections
develop with appropriate signals to the brain (received in appropriate 
time), and it takes maximum time (about 26 to 30 years) for human. Compared
to this, the animal brain development (development of their
inter-neuron connections) takes very little time and in fact it
ceases almost immediately after their birth. It appears therefore that 
I differ from great artists, scientists or mathematians, mostly in our
respective developments (of synaptic connections) after our births! 

As mentioned already, neurons are electrical devices and can be in 
two functional states: firing state (if the aggregate synaptic voltage 
in the cell exceeds the threshold) or quiescent  state (otherwise).
Neurophysiologists McCullogh and Pitts (see e.g., [5]) therefore 
proposed the idea of functional modelling of a single neuron by a two
state device like an electrical valve or an electronic transistor. The
consequent excitement came from the realization that the present day
computer workstations already employ about $10^8$ transistors (comparable
to the neuron number in a pegion's head) and the transistors, being
electronic systems, work much faster (typical time scale being $10^{-8}$
seconds) while the ionic flow rate in the neurons are much slower
(with typical time scale of the order of $10^{-1}$ to $10^{-3}$ seconds).
Once the inter-neuronal connection architecture in the brain is understood,
its artificial implementation on a silicon device may become extremely 
powerful!

As mentioned before, although we know now a little bit about the structure
and function of a single neuron, we are still in the dark about the growth
of the neural network through the inter-neuron synaptic connections. We
do not know yet the precise algorithms followed during the learning processes
to develop these connections. 

One can use a digital or binary representation of any pattern using pixel
decompositions. Each such pattern can then be made an `attractor' configuration
of the network (of 
binary neurons) following  a network dynamics. Starting from
any `corrupted' or distorted version of that pattern then the dynamics
of the network brings back the `learned' pattern as the dynamics get attracted
towards that. The dynamical matrix elements, representing the synaptic
interaction between the neurons, depend on the pattern the network intends
to remember or get attracted to. Two independent patterns then demand 
differently for these matrix elements. Memory of a large number of such
patterns then demand conflicting or frustrating requirements for the synaptic
connections or the matrix elements. This is a generic feature for such networks.
In fact, this frustration leads to a macroscopic number of local attractors
of the dynamics of the network, which helps large memory size etc;
without frustration, the network would have only two attractors (and hence
two memory states). In the Hopfield model, one defines an energy function 
in the pattern configuration space such that the learned patterns correspond to
local energy minima, whereas the corrupted or distorted patterns correspond to
higher energies. Any dissipative energy minimisation dynamics then brings
the system to the local minima or memory state if the starting configuration
was within its domain of attraction. In this model, the synaptic connections 
are taken, following Hebbs, symmetric and its magnitude given by the algebric
sum of the inter-neoron interactions required for each of the pattern to be
learned or memorised. The resultant interactions then become random not only
in magnitude but also in sign. This frustration leads to a maximum memory
size of the network (capable of recalling from distorted patterns) about 14\%
of the network size, given by the number of neurons in the 
network. The network gets confused
if more patterns are put in it! In the Rosenblatt perceptron model, these
dynamical matrix elements (synaptic connections) evolve dynamically by
minimising errors in predicting the `unseen' part of the pattern. After
some initial `supervision', such networks perform various pattern recognition
jobs satisfactorily (see e.g., [5]). 

\medskip

\noindent {\it Criticsms:} As mentioned in the previous section, although
there have been intreguing developments and consequent excitements 
regarding the possibility of artificial intelligence and mind, as good
as  those of the human, severe criticsms of such Strong AI hypothesis  have
been forwarded by several scientists. The Strong AI states (cf. Crick [6])
that a ``Computer will not only have mental states as its emergent
property, the implemented program will by itself constitute the
mind". Penrose [3] argues that such a 
machine can not have consciousness (cf. emperor's new cloth) which in
his view is complicated by
 quantum mechanical entanglements. Chalmers [7] argues that
even if such a machine performs all these (computations and pattern
recognitions), it can not be `self-conscious'.
The argument is that a computer program is defined purely syntactically,
and that the syntax itself is not enough to guarantee the presence of mind. 
My stomach pain is my personal feeling and I am conscious of that, while
the physiological disorder and neurological processes following that
are objective facts for a physician identifying the cause of my pain;
they are not identical. Searle [8] developed a `chinese room' argument 
to refute the Strong AI hypothesis. The argument runs as follows: Even 
if I do not understand chinese, I can behave like a chinese by following
a set of preassigned (say translated) rules or programs. Within these set of 
rules (program) I will appear to behave as understanding it; although I 
do not! Thus (a) programs are entirely syntactical, (b) minds have
semantics and (c) syntax is not the same as, nor by itself sufficient, for
semantics. This three step chinese room argument therefore proves 
``Programs are not minds" (cf. [8]).

\bigskip

\noindent {\bf IV. Indian Concept of mind \& Bose's nervous mechanism
of plants:}

\medskip

\noindent In Upanishad (1500 B.C. - 1000 B.C.), the mind was argued
to be composed of the heart and the brain. In fact, in Pra\'sna, one 
gets even a description of the physiological structure of the mind [9].
From the heart, 101 `na\-di' or `dhamani' gets out, each of which apparently
branches out in 100 thinner and tinier 
branches, and so on. It says, in total about 72000
`nadi' or `dhamani' are spread all throughout our body and the brain. 
Proper function of our brain depends on all of them [9]. This 3000 year
old crude and speculative model
 might be compared with our present (established) knowledge of 
about $10^{12}$ neurons in the human brain! Upanishad then argues that
the external objects or processes then induce, through the senses and
conveyed through the `nadi's, `Manas' (or sense-data) which in turn induces
`Buddhi' (becomes unmanifest) giving rise to `Purusa-Atman' (self or
ego) which finally melts down to `Chitta' (consciousness). These ideas
about the consciousness of our mind were later evolved in 
Bhagavad Gita (300 B. C.)
and in particular in Buddhism (273 B. C. - 200 B. C.) [10].

The  idea or the philosophical doctrine that the mind is not essentially
confined only to a  small 
part part of the body (for example the brain), and that
it disperses all over the body seemed to be a dominating one in ancient Indian
thoughts. In fact, even the treatment and control of the mental
processes, as advocated and prescribed in Susrut (500
B. C.) and in Yoga and Tantraloka (cf. [10]), involved some thoughtful
and thorough exercise of the various parts of our body! It is indeed
unfortunate that scholastic follow-ups, scientific investigations following
these ideas and their refinements are nonexistant or insignificant. Even
documents and books on these developments are scarce (cf. [9, 10]).

It is particularly heartening to `discover' in this context, the
experimental work of Jagadish Bose in the last century on the nervous
mechanism of plants [11]. It is well known, plants do not have brains,
and hence do not have any neuronal cells or their network like us. Yet,
the plants do indeed perform computations for adjusting and responding
to the changing environments. Plants do these calculations slowly, but
surely. Imagine the response, say within a week, of a plant in a 
suddenly darkened area with sunlight coming only from an angle, or
take the case of a creeper plant climbing up a window grill or a pillar
with its tentacles or branches! Imagine the amount of computations
involved in `recognising' the structure of the
neighbouring posts or grilles, in `finding'  
 their minimum cross-sections and in holding them by growing
around the necks of the neighbouring structures. Do they also have
personal feelings? Are they self-conscious? We do not know.

Through his pioneering experiments, J. C. Bose [11] showed about 
hundred years ago that the plant cells are excitable and can transmit 
millivolt order electrical signals at about 10-40 millimeter per second
speed. 
Through these electrical signals, these cells communicate  in 
coordinating their responses to the environment (see Fig. 3). This
analysis and `recognition' of the changes in the external 
environment is therefore performed by the plants, according to Bose, 
through its extended (nervous) cellular network all across its 
trunks, branches and leaves. It may be mentioned that this partial
electrical signalling between the plant cells, like those in
the neurons of the animals, is now a fairly established fact; although,
 for a long  period after Bose's pioneering work, the plant physiologists
did not accept it and considered the inter-cell signalling to be 
purely chemical diffusion in origin (see e.g., Shephard [11]). This 
observation of extended computation or processing of the environmental
signals all over the living body of the plant or the animal is in
fact very much in conformity with the ancient Indian idea of mind-body
relationship. Again, not much development has taken place in this direction. 

\bigskip

\noindent {\bf Concluding remarks:}

\medskip

\noindent Present analysis of the brain structure and of the
neural computation process indicates  how a collective computing property
(like consciousness) might 
emerge out of a network of (about $10^{12}$) neurons or transistors. 
Unlike the present day computers, our brain calculates in a distributed way. 
It employs parallel processing involving
 almost all the neurons in the brain for
each computing operation.
The observation by Bose [11] on the plant nervous system 
suggests that computations can
be much more distributed than we can think today. Plants do not have any
brain and yet they compute using their cells all over the plant body.
Can individual brains interact? Electrical contacts or interactions
are not possible, but perhaps socially? World population today is about 
$10^{10}$, and every one of us has got a brain to perform simple tasks. 
Is it possible that collective computational
capacity of many such  brains
might give rise to 
higher order comuptational abilities and
the (social) consciousness we are so familiar with? Isolated
person like Robinson Crusoe's  brain may not 
generate it. But an interactively evolving society perhaps
develops it necessarily?  Partial or  indifferent participation 
may then lead to different perceptions, `value judgements' and ethics.
Compared to the innate history of the universe, human histroy therefore becomes
accessible to value judgements and conscious evaluation (cf. [12]).
Such a possiblity seems to be pretty close to the ideas 
floated by the major Indian schools of thought, starting from
Upanishad. If this is true, machines can 
also have such consciousness; only perhaps collectively!

\bigskip

\noindent {\bf Acknowledgement:} I am grateful to A. K. Bag, R. Banerjee,
P. Bhattacharyya, R. L. Brahmachary, A. Chatterjee, A. Dutta, A. Kundu, 
S. Pradhan and P. Mitra 
for several useful and encouraging comments.

\vskip 1 cm

\noindent {\bf References:}

\medskip

\noindent 
[1] A. N. Whitehead and B. Russell, {\it Principia Mathematica}, Vols. I-III,
Cambridge Univ. Press, Cambridge (1962)

\medskip

\noindent
\noindent
[2] G. T. Kneebone, {\it Mathematical Logic \& Foundations of Mathematics},
Van Nostrand, London (1963)

\medskip

\noindent
[3] R. Penrose, {\it Emperor's New Mind}, Oxford Univ. Press, Oxford (1989)

\medskip

\noindent
[4] M. J. Feigenbaum, {\it Journal of Statistical Physics} {\bf 19} (1978) 25
; E. Witten in {\it Critical Problems in Physics}, Eds. V. L. Fitch et al,
Princeton Univ. Press, Princeton (1997) pp. 271-280

\medskip

\noindent
[5] D. J. Amit, {\it Modelling Brain Functions}, Cambridge University Press,
Cambridge (1989)
 
\medskip

\noindent
[6] F. Crick, {\it The Astonishing Hypothesis: The Scientific Search for
the Soul}, Simon \& Schuster, London (1994); see also P. Churchland,
{\it The Engine of Reason, The Seat of the Soul}, MIT press, Massachusetts
(1995)

\medskip

\noindent
[7] D. J. Chalmers, {\it The Conscious Mind}, Oxford Univ. Press, Oxford (1996)

\medskip

\noindent
[8] J. R. Searle, {\it The Mystery of Consciousness}, Granata Books, London
(1997)

\medskip

\noindent
[9] P. T. Raju, in {\it The Cultural Heritage of India}, Vol. III, Ramakrishna
Mission Institute of Culture, Kolkata (1983) pp. 507-519, 581-607

\medskip

\noindent
[10] R. L. Gregory (Ed.), {\it The Oxford Companion to Mind}, Oxford 
Univ. Press, Oxford (1998) pp. 146-47, 357-61

\medskip

\noindent
[11] J. C. Bose, {\it The Nervous Mechanisms of Plants}, Longmans, London
(1926); See also V. A. Shepherd, {\it Current Science} {\bf 77} (1999) pp.
189-195

\medskip

\noindent [12] I. Berlin, {\it The Proper Study of Mankind}, Pimlico,
London (1998) pp. 17-58.

\vskip 1 cm

\noindent {\bf Figure captions:}

\medskip

\noindent Fig. 1. Schematic structure of a neuron.

\medskip

\noindent Fig. 2. Development of inter-neuron synaptic connections in
the human visual cortex after the birth: (from left to right) newborn,
three month old and two year old infant [From T. H. Bullock, R. Orkand
and A. Grinnel, {\it Introduction to Nervous Systems}, Freeman, San Francisco
(1977)].

\medskip

\noindent Fig. 3. Electrical pulsations in Desmodium, measured by inserting
the probe slowly (0.1 mm per turn) within the tissues [from Bose [11]].
 
\newpage
\begin{figure}
\centerline{\psfig{file=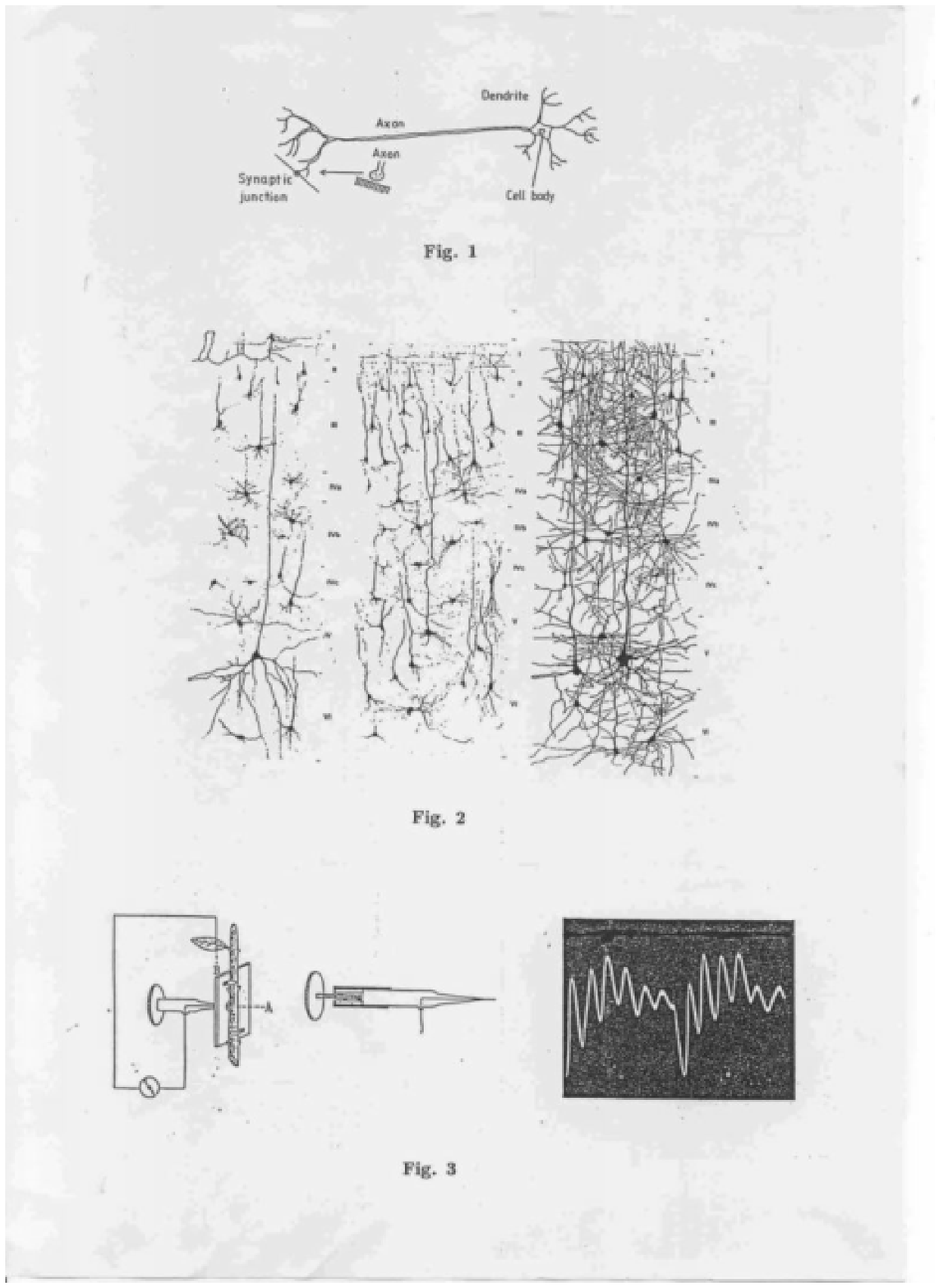}}
\end{figure}
\end{document}